\documentclass[prl,twocolumn,showpacs,preprintnumbers]{revtex4-1}
\usepackage{graphicx}
\usepackage{amsmath}
\usepackage{amssymb}
\usepackage{times}
\usepackage{color}
\usepackage{ulem}

\def\D{\Delta}
\def\d{\delta}

\def\g{\gamma}

\def\w{\omega}

\def\Vec#1{\mathbf #1}
\begin{document}

\title{Exploring the dark side of cuprate superconductors: $s$-wave symmetry of the pseudogap}

\author{S. Sakai$^{1,2,3}$, S. Blanc$^4$, M. Civelli$^5$, Y. Gallais$^4$, M. Cazayous$^4$, M.-A. M\'easson$^4$, J. S. Wen$^6$, Z. J. Xu$^6$, G. D. Gu$^6$, G. Sangiovanni$^{7,8}$, Y. Motome$^2$, K. Held$^8$, A. Sacuto$^4$, A. Georges$^{1,3,9,10}$, and M. Imada$^{2,3}$}

\affiliation{$^1$Centre de Physique Th\'eorique, \'Ecole Polytechnique, CNRS, 91128 Palaiseau, France,\\
$^2$Department of Applied Physics, University of Tokyo, Hongo, Tokyo 113-8656, Japan,\\
$^3$JST-CREST, Hongo, Bunkyo-ku, Tokyo, 113-8656, Japan,\\
$^4$Laboratoire Mat\'eriaux et Ph\'enom$\grave{e}$nes Quantiques (UMR 7162 CNRS), Universit\'e Paris Diderot-Paris 7, Bat. Condorcet, 75205 Paris Cedex 13, France,\\
$^5$Laboratoire de Physique des Solides, Universit\'e Paris-Sud, CNRS, UMR 8502, F-91405 Orsay Cedex, France,\\
$^6$Matter Physics and Materials Science, Brookhaven National Laboratory (BNL), Upton, NY 11973, USA,\\
$^7$Institut f\"ur Theoretische Physik und Astrophysik, Universit\"at W\"urzburg, Am Hubland, D-97074 W\"urzburg, Germany,\\
$^8$Institute for Solid State Physics, Vienna University of Technology, 1040 Vienna, Austria,\\
$^9$Coll\`ege de France, 11 place Marcelin Berthelot, 75005 Paris, France,\\
$^{10}$DPMC, Universit\'e de Gen\`eve, 24 quai Ernest Ansermet, CH-1211 Gen\`eve, Suisse
}
\date{\today}
\begin{abstract}
We reveal the full energy-momentum structure of the pseudogap of underdoped high-$T_\text{c}$ cuprate superconductors. Our combined theoretical and experimental analysis explains the spectral-weight suppression observed in the B$_\text{2g}$ Raman response at finite energies in terms of a pseudogap appearing in the single-electron excitation spectra above the Fermi level in the nodal direction of momentum space.
This result suggests an $s$-wave pseudogap (which never closes in the energy-momentum space), distinct from the $d$-wave superconducting gap.
Recent tunneling and photoemission experiments on underdoped cuprates also find a natural explanation within the $s$-wave-pseudogap scenario.
\end{abstract}
\maketitle

The superconducting gap of high-$T_\text{c}$ cuprate superconductors has a $d$-wave symmetry\cite{tsuei97} with zero-point nodes in momentum space, in contrast to the nodeless $s$-wave gap of conventional superconductors. In the underdoped regime another gap, called a pseudogap, exists even above the critical temperature $T_\text{c}$. 
The relation between pseudo- and superconducting gaps has been a controversial issue whose understanding may provide long-sought insights into the mechanism of high-temperature superconductivity\cite{norman05,hufner08,millis06,kanigel08,yang08,hashimoto10,he11,hoffman10}. According to a broad class of theories, the pseudogap is a continuation of the superconducting gap into a regime of incoherent Cooper pairs. A competing class of theories holds instead that the pseudogap is a manifestation of a new instability; therefore, it should be different from the superconducting gap. Most of these theories \cite{anderson87,emery95,chakravarty01,balents98,franz01,yang06} assume a $d$-wave structure of pseudogap, since angle-resolved photoemission spectroscopy (ARPES) \cite{damascelli03} has observed the pseudogap vanishing in the nodal region in a fashion reminiscent of a $d$-wave superconducting gap. Nevertheless ARPES can only access the occupied side of the electronic spectra. Therefore the determination of the complete structure of the pseudogap is an essential ingredient missing in order to unveil the real connection between pseudo- and superconducting gap.

In this Letter, we explore the pseudogap structure on the ``dark" (unoccupied) side of underdoped cuprates, by combining cellular dynamical mean-field theory (CDMFT)\cite{kotliar01} with Raman-spectroscopy experiments.
In stark contrast with the assumption of a $d$-wave pseudogap, we find an $s$-wave pseudogap which accounts for various anomalous properties of the nodal (B$_\text{2g}$) and antinodal (B$_\text{1g}$) Raman responses, as well as of ARPES and scanning tunneling microscopy (STM).
Our study focuses on the pseudogap phase above $T_\text{c}$, leaving open the question about possible competing orders under the superconducting dome, where quantum oscillation experiments on YBa$_2$Cu$_3$O$_{6.5}$ \cite{doiron07,sebastian08} have revealed a dramatic reconstruction of the electronic structure.

\begin{figure}[htb]
\center{
\includegraphics[width=0.5\textwidth]{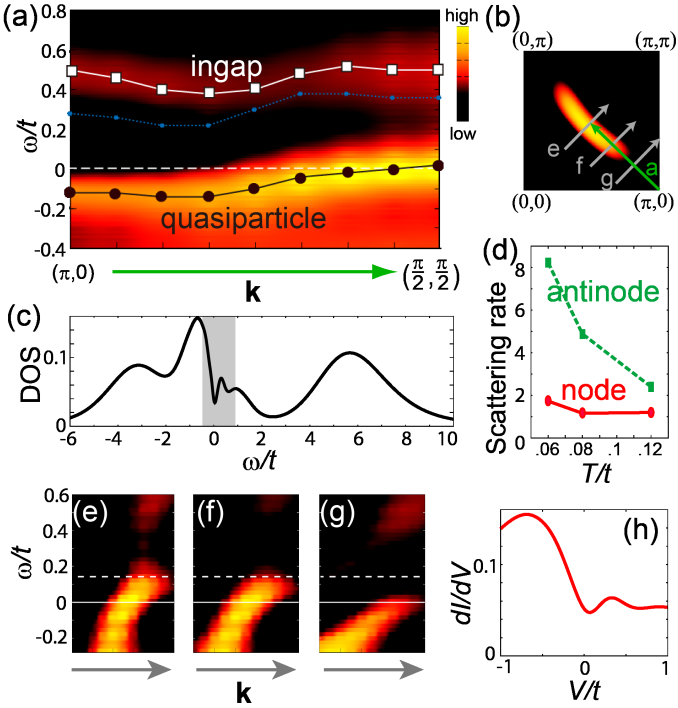}}
\caption{Theoretical spectral intensity at $T$$=$0.06$t$ and $p_\text{th}$$=$0.05 (a) in the energy-momentum space and (b) in the first quadrant of the Brillouin zone at low energy ($\w$$\sim$0). Black circles and white squares denote, respectively, the peak positions of quasiparticle and in-gap bands, which are separated by an $s$-wave pseudogap. The dotted blue curve plots the position of the maximal-scattering rate (i.e., the imaginary part of self-energy) at each momentum. The green arrow in (b) denotes the momentum cut used in (a).
(c) Local DOS in a wide energy range displaying the Hubbard bands ($\w$$<$0 and $\w$$\sim$6$t$). The shaded area denotes the low-energy region plotted in (a). 
(d) Temperature dependence of the scattering rate within the pseudogap at $\w$$=$0.36$t$ (node) and 0.28$t$ (antinode). (e-g) Spectral function along the cuts depicted in (b), comparable to available ARPES data\cite{yang08}. The dashed line indicates the upper energy limit reached in Fig.~3{\bf e-g} of Ref.~\onlinecite{yang08}. (h) Tunneling conductance at $T$$=$0.06$t$, comparable to e.g., Fig.~5{\bf B}(inset) of Ref.~\onlinecite{pushp09}.
}
\label{fig:peak}
\end{figure}

Raman B$_\text{2g}$ (B$_\text{1g}$) spectroscopy, obtained from cross polarizations along (45$^\circ$ from) the Cu-O bonds \cite{devereaux94}, has been performed on an underdoped Bi$_2$Sr$_2$CaCuO$_{8+\d}$ (Bi2212) single crystal, grown with a floating zone method ($T_\text{c}$$=$74K). The hole-doping concentration $p$$\sim$0.11 has been achieved by changing the oxygen content only. A triple grating spectrometer (JY-T64000) equipped with a nitrogen cooled CCD detector was used. 
All the measurements have been corrected for the Bose factor and the instrumental spectral response, and are thus proportional to the imaginary part of the Raman response function. 

\begin{figure*}[tb!!]
\center{
\includegraphics[width=0.7\textwidth]{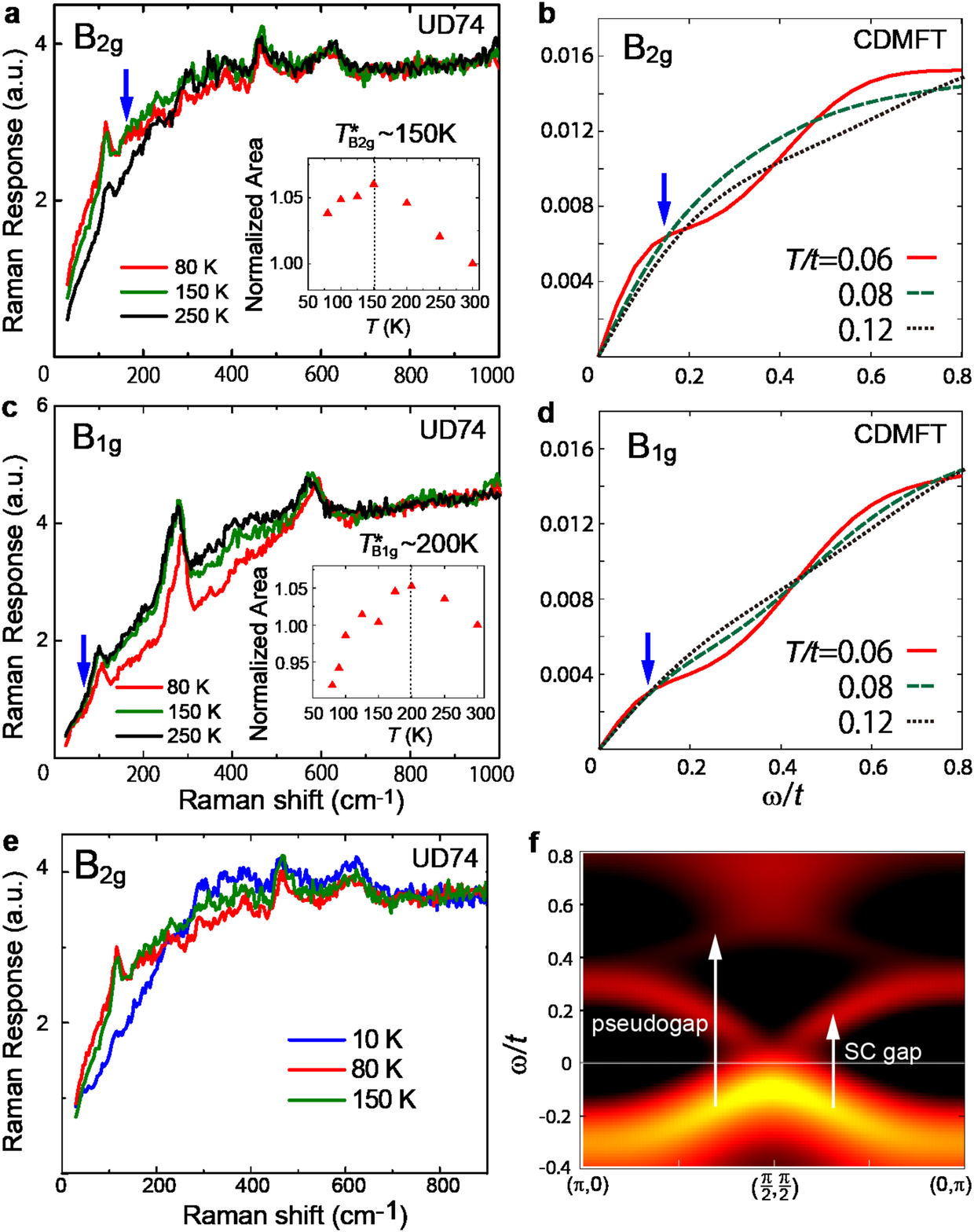}}
\caption{B$_\text{2g}$ Raman spectra obtained by experiments (a) on an underdoped Bi2212 ($T_\text{c}$$\sim$74K, $p$$\sim$0.11) and by the CDMFT (b) in the underdoped regime ($p_\text{th}$$=$0.05).
(c),(d) The same for B$_\text{1g}$ spectra. The blue arrows mark the starting energy points of pseudogap depression. The insets in (a) and (c) plot the integrated Raman weight normalized at $T$$=$300K as a function of temperature. 
(e) Experimental B$_\text{2g}$ Raman response in the normal ($T$$=$150K and 80K) and superconducting ($T$$=$10K) states. 
(f) Theoretical spectral intensity calculated with CDMFT + Exact diagonalization method for a 2$\times$2 cluster in the superconducting state.
The white arrows denote excitations beyond the pseudo- and superconducting (SC) gaps in the nodal region, which contribute to the B$_\text{2g}$ Raman intensity.}
\label{fig:raman}
\end{figure*}

For the theoretical analysis we have adopted a minimal model of the Cu-O planes: The two-dimensional Hubbard model with the (next-)nearest-neighbor transfer integral $t$($t'$$=$-0.2$t$) and the onsite Coulomb repulsion $U$$=$$8t$. CDMFT has been implemented on a 16-site cluster\cite{sakai12}, which is much larger than a 4-site cluster previously used \cite{sakai09,sakai10,senechal04,civelli05,kyung06,stanescu06,liebsch09,civelli08,civelli09} and is solved with the continuous-time quantum Monte Carlo (CTQMC) method\cite{gull11} (except for the case of Fig.~\ref{fig:raman}(f), where the exact diagonalization method for a 2$\times$2 cluster was employed). The momentum-dependent quantities have been extracted from the cluster by using the cumulant interpolation scheme\cite{stanescu06} (see Refs.~\onlinecite{sakai12} and \onlinecite{civelli09}).
 
In order to avoid a severe sign problem in the CTQMC, we are forced to adopt a theoretical doping $p_\text{th}$$=$0.05 smaller than the experimetal $p$$\sim$0.11.
However, our goal is to identify general doping-independent properties of the pseudogap phase rather than to simulate a specific material at a specific doping. We shall therefore trace a qualitative comparison between theory and experiments, and in order to facilitate it we (i) provide an order-of-magnitude value by setting the energy scale of our theory $t$$=$0.3eV \cite{andersen95} (1eV$\sim$1.2$\times$10$^4$K$\sim$8.1$\times$10$^3$cm$^{-1}$) and (ii) divide by a factor of 1.5 the theoretical energy scales directly related to the pseudogap amplitude, because the pseudogap energy scale at 5\% doping is about 1.5 times larger than that at 11\%\cite{damascelli03,devereaux07}.

Figure \ref{fig:peak}(a) presents the theoretical spectral weight $A(\Vec{k},\w)$ plotted along the $(\pi,0)$-$(\pi/2,\pi/2)$ $\Vec{k$}-cut [labeled ``a" in Fig.~\ref{fig:peak}(b)]. 
Here we focus on the low-energy region shaded on the local density of states (DOS) in Fig.~\ref{fig:peak}(c). We find a coherent dispersing quasiparticle band (black circles) crossing the Fermi level ($\w$$=$0) in the nodal region $\Vec{k}$$\sim$$(\pi/2,\pi/2)$, and a less dispersive and less coherent band (white squares) at positive energies $\sim$0.5$t$. We call this latter an in-gap band because it arises inside the Mott gap [the wide 0$<$$\w$$<$4$t$ region in Fig.~\ref{fig:peak}(c)]. Between these two bands a gap opens, which we identify with the pseudogap observed in cuprates. The most striking feature is that our pseudogap never vanishes in the energy-momentum space, even in the nodal direction. Hence, differently from the $d$-wave superconducting gap, it has an $s$-wave symmetry\cite{sakai09,sakai10} with no node. This structure nevertheless looks like a $d$ wave if observed in the negative-energy plane, because the nodal region is gapless below the Fermi level. Several numerical calculations \cite{kyung06,tohyama04,sakai09,sakai10} on smaller clusters without any {\it a priori} assumption have also indicated a similar pseudogap structure.
In our theory the pseudogap is a pure outcome of the parent Mott insulator\cite{anderson87}. 
The appearance of a strong-scattering surface in the energy-momentum space [dotted blue line in Fig.~\ref{fig:peak}(a)] drives the metallic system into the Mott insulator by suppressing spectral weight\cite{senechal04,sakai09,stanescu06,yang06,civelli05,berthod06,yamaji11}. As this surface is closer to the Fermi level in the antinodal than in the nodal region\cite{sakai09}, a (pseudo)gap in the spectra opens around $\w$$=$0 at the antinodes while it shifts to positive energies in the nodal region\cite{sakai09,kyung06,stanescu06}, where a Fermi arc is observed [Fig.~\ref{fig:peak}(b)].
 
Experimentally there are few studies on the momentum structure of the unoccupied spectra. ARPES at a relatively high temperature can detect spectra slightly above the Fermi level by analyzing thermally-populated states. For an underdoped Bi2212 sample ($T_\text{c}$$=$65K, $p$$\sim$0.09) at $T$$=$140K, Yang {\it et al.}\cite{yang08} obtained the unoccupied spectra below 0.04eV. 
These results (reproduced in Figs.~S1e-g) are in good agreement with our theoretical spectra, Figs.~\ref{fig:peak}(e)-(g).
A dashed white line in each panel indicates the positive energy-window which we estimate was accessed in the ARPES experiment. In the nodal direction [Fig.~\ref{fig:peak}(e)] the pseudogap opens above the window and the electronic dispersion close to the Fermi level appears rather symmetric. In moving to the antinodal region [Fig.~\ref{fig:peak}(g)] the pseudogap shifts down into the energy window and the electronic dispersion displays a marked electron-hole asymmetry. 
This again evidences a radical difference between pseudogap and superconducting gap, whose hallmark is represented by two particle-hole-symmetric Bogoliubov bands.

The tunneling conductance $dI/dV$, albeit missing momentum resolution, can also provide valuable information on the unoccupied spectra. In Fig.~\ref{fig:peak}(h) we plot the theoretical tunneling conductance calculated by $\frac{dI}{dV}=-\int d\w f'(\w-eV)D(\w)$, where $f'$ is the energy derivative of the Fermi distribution function $f$, $-e$ is the electron charge, and $D(\w)$ is the cluster density of states.
First of all, the $dI/dV$ curve has a smaller weight on the unoccupied side, reflecting the underlying projective nature of strong correlation \cite{anderson-ong06}. The hump around $V$$=$$0.4t$ reflects the  ingap band at positive energies seen in Fig.~\ref{fig:peak}(a). 
The $s$-wave pseudogap predominantly opening on the unoccupied side elucidates a further asymmetric shape in the $dI/dV$ curve, which is observed by STM for strongly underdoped samples (see e.g., the inset in Fig.~5{\bf B} of Ref.~\onlinecite{pushp09}).
For a more quantitative comparison with ARPES and STM results, see Sections.~II and IV in the Supplementary Information.

However, ARPES\cite{yang08} and STM\cite{pushp09} are limited, respectively in energy range and in momentum resolution.
Hence key information about the presence of a gap in the {\it unoccupied} spectra in the {\it nodal} region is still missing in experiments.
We have therefore performed Raman spectroscopy\cite{devereaux07}, which, albeit in a less direct way, can separately access the nodal (B$_\text{2g}$) and antinodal (B$_\text{1g}$) electronic structures, as well as the wide energy region above the Fermi level. 

Theoretical Raman spectra have been calculated within the bubble approximation from the CDMFT single-particle spectra
\begin{align}
\chi_\mu''(\w)=2\int \frac{d\Vec{k}}{(2\pi)^2} \g_\mu^2(\Vec{k}) 
 \int_{-\infty}^\infty &d\w' A(\Vec{k},\w')A(\Vec{k},\w+\w')\nonumber\\
\times &[f(\w')-f(\w+\w')]
\label{eq:raman}
\end{align}
with $\g_{\rm B1g}$$=$$\frac{1}{2}[\cos(k_x)-\cos(k_y)]$ and $\g_{\rm B2g}$$=$$\sin(k_x)\sin(k_y)$. This is known to give a reasonable estimate for B$_\text{2g}$ response \cite{devereaux99}, on which our main result of $s$-wave pseudogap relies. Vertex corrections can be more significant in B$_\text{1g}$ geometry. However a recent study \cite{lin12} based on the dynamical cluster approximation \cite{maier05} shows that the corrections are still small in a low-energy region where the antinodal pseudogap opens. These considerations together with the nice correspondence with the experimental results (as we will show in the following) support our theoretical analysis.

We first point out that the CDMFT Raman spectra [Figs.~\ref{fig:raman}(b),(d)] well reproduce the rather broad incoherent electronic response observed in experiments [Figs.~\ref{fig:raman}(a),(c)]. This broad feature is the outcome of mixing low-energy coherent quasiparticle excitations with incoherent high-energy ones (e.g., the Hubbard bands), which are captured within CDMFT. This description would not be possible by approaches employing only low-energy quasiparticles.

Second, in the B$_\text{2g}$ (nodal) response the slope at $\w$$=$0, which is proportional to the quasiparticle lifetime, increases with lowering temperature, both in experiments [Fig.~\ref{fig:raman}(a)] and in theory [Fig.~\ref{fig:raman}(b)]\cite{phonon}.
This behavior is consistent with a metallic Fermi arc observed around the node by ARPES\cite{damascelli03} and within CDMFT [Fig.~\ref{fig:peak}(b)]\cite{sakai09,sakai10,sakai12,civelli05,kyung06,stanescu06,liebsch09,macridin06}. 
The low-energy slope of the B$_\text{1g}$ (antinodal) response shows instead little temperature dependence [Figs.~\ref{fig:raman}(c),(d)]\cite{devereaux07,venturini02,gallais05,opel00}. This signals a nonmetallic behavior at the antinodes, where the pseudogap indeed opens at the Fermi level\cite{sakai09,sakai10,kyung06,stanescu06,liebsch09,lin10,macridin06,ferrero09}.

We now look at the behavior of the B$_\text{2g}$ [B$_\text{1g}$] response in the intermediate-energy interval (0.15-0.45$t$ [0.1-0.45$t$] in CDMFT and 150-600 [50-600]cm$^{-1}$ in experiments), whose onset is indicated by the blue arrow in Figs.~\ref{fig:raman}(a)-(d). 
In this energy range a depression should result from the appearance of a pseudogap. 
The nontrivial fact is that this depression takes place not only in the B$_\text{1g}$ symmetry, where one expects to see the antinodal pseudogap, but also in the B$_\text{2g}$ symmetry. 
The interval of the depression in the B$_\text{2g}$ [B$_\text{1g}$] theoretical Raman response is 0.3$t$$\sim$480cm$^{-1}$ [0.35$t$$\sim$560cm$^{-1}$] wide (taking into account the above-mentioned factor 1.5 due to the difference between $p$ and $p_\text{th}$), in a good agreement with the experimental value, 450 [550]cm$^{-1}$.
Notice that the energy endpoint of the B$_\text{2g}$ depression is nearly equal to that of B$_\text{1g}$; 600cm$^{-1}$ in experiment and 0.45$t$ in theory. A similar depression in B$_\text{2g}$ response was previously reported for other underdoped cuprates\cite{gallais05,opel00,nemetschek97}, where it was attributed to the gap opening at the Fermi level away from the node. However, our study reveals a novel mechanism due to a gap above the Fermi level in the nodal region, as we shall explain below.

Another remarkable property is that the antinodal ($T^\ast_\text{B1g}$) and nodal ($T^\ast_\text{B2g}$) pseudogap-crossover temperatures are different. In the insets of Figs.~\ref{fig:raman}(a),(c) we plot the area under the electronic response (up to 800cm$^{-1}$) as a function of temperature. The maxima provide an estimation of the pseudogap-crossover temperature $T^\ast_\text{B2g}$$\sim$150K and $T^\ast_\text{B1g}$$\sim$200K. In the CDMFT result for the intermediate energies, while the B$_\text{1g}$ response monotonically decreases from $T$$=$0.12$t$ to 0.06$t$ [Fig.~\ref{fig:raman}(d)], the B$_\text{2g}$ response is not monotonic: It increases from $T$$=$0.12$t$ to 0.08$t$ and then decreases by further lowering the temperature to $T$$=$0.06$t$ [Fig.~\ref{fig:raman}(b)].
Thus, we find two different crossover temperatures in the CDMFT results, too; $T^\ast_\text{B2g}$$\sim$0.08$t$$\sim$180K and $T^\ast_\text{B1g}$$\gtrsim$0.12$t$$\sim$280K (again taking into account the factor 1.5), still in reasonable agreement with the experimental values.
We summarize these comparisons of energy scales in Section I of the Supplementary Information.

\begin{figure}[thb]
\center{
\includegraphics[width=0.4\textwidth]{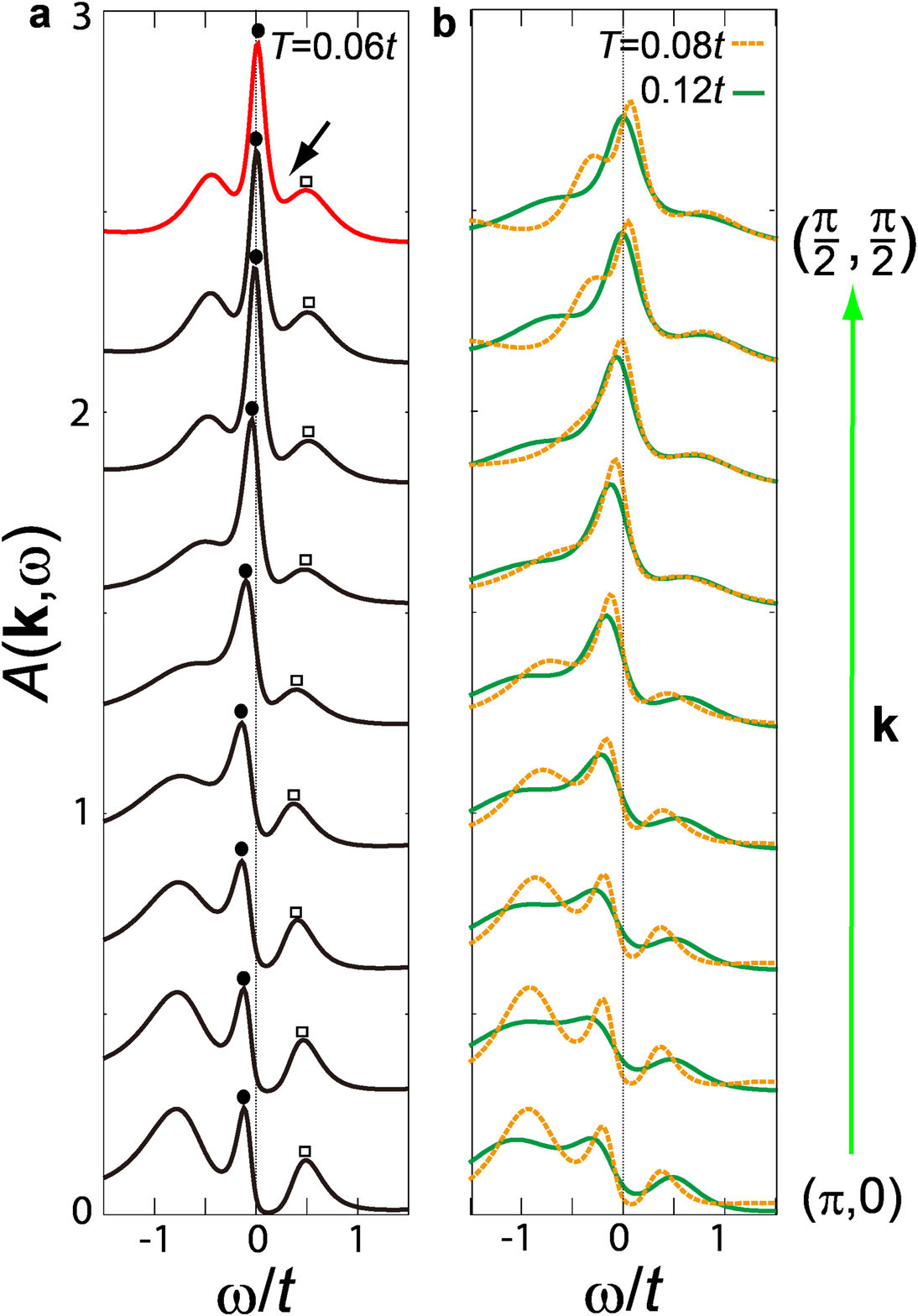}}
\caption{Energy distribution curves of the single-particle spectra along the $(\pi,0)-(\pi/2,\pi/2)$ line (a) at $T$$=$0.06$t$ and (b) at $T$$=$0.08$t$ and 0.12$t$. The black circles (white squares) denote the quasiparticle (in-gap) peak plotted in Fig.~\ref{fig:peak}(a). The arrow at the top curve denotes the pseudogap. For clarity the curves are offset by 0.3.}
\label{fig:kdep}
\end{figure}

We now analyze the Raman response in terms of the CDMFT spectra $A(\Vec{k},\w)$. In Fig.~\ref{fig:kdep}(a), sandwiched in between the quasiparticle peak (black bullets) close to the Fermi level and an ingap peak (white squares) at $\w$$\sim$0.5$t$, a depression (arrow) smoothly continues from the antinode to the node. This has been identified with the pseudogap in Fig.~\ref{fig:peak}(a). 
This pseudogap depression in $A(\Vec{k},\w)$ originates the depression in the Raman responses [see  Eq.~(\ref{eq:raman})], as indeed seen in Figs.~\ref{fig:raman}(b) and (d). In particular, the presence of the pseudogap at positive energy in the nodal region leads to the depression in the B$_\text{2g}$ Raman response in the intermediate-energy interval. In Fig.~\ref{fig:kdep}(b) the pseudogap persists up to $T$$=$0.12$t$ around the antinode, while it is almost lost at $T$$=$0.08$t$ around the node, consistently with the difference between $T^\ast_\text{B2g}$ and $T^\ast_\text{B1g}$ observed in the Raman spectra. The difference originates from different temperature dependences of scattering rates in the nodal and antinodal regions, as extensively reported in experiments \cite{damascelli03,opel00,devereaux07}, and as shown in Fig.~\ref{fig:peak}(d), where the CDMFT nodal and antinodal maximal-scattering rates within the pseudogap are plotted against temperature.

We finally turn to the experimental Raman response in the superconducting state and show that it also supports the $s$-wave pseudogap. Figure \ref{fig:raman}(e) compares the B$_\text{2g}$ responses below and above $T_\text{c}$. In general, upon the opening of a superconducting gap below $T_\text{c}$, spectral weight is removed from the Fermi level. This is true also for a $d$-wave superconducting gap around the nodal point. Accordingly, the B$_\text{2g}$ Raman response decreases at low energy ($\w$$<$200cm$^{-1}$) and increases at higher energy (200cm$^{-1}$$<$$\w$$<$700cm$^{-1}$). 
Interestingly, the latter increase emerges mostly within the pseudogap energy range (150cm$^{-1}$$<$$\w$$<$600cm$^{-1}$).
A similar behavior was reported also in YBa$_2$Cu$_3$O$_{7-x}$, Bi$_2$Sr$_2$(Ca$_{0.62}$Y$_{0.38}$)Cu$_2$O$_{8+\d}$\cite{nemetschek97,opel00}, and HgBa$_2$CuO$_{4+\d}$\cite{gallais05}, showing that it is common to various cuprates. 
This suggests that the superconducting gap is substantially smaller than the pseudogap, particularly in the nodal region. Namely, the Bogoliubov quasiparticle bands emerge below $T_\text{c}$ in the nodal region inside the pseudogap energy range.
This scenario is illustrated in the explanatory figure \ref{fig:raman}(f), which plots a 2$\times$2 CDMFT $A(\Vec{k},\w)$ in the superconducting state \cite{civelli09} along the momentum-space cut ($\pi$,0)-($\pi/2$,$\pi/2$)-(0,$\pi$). Here we use a simple $\cos k_x - \cos k_y$ form (which is widely supported in experiments\cite{damascelli03}) for interpolating the CDMFT $d$-wave superconducting gap.
In particular, around the node, the superconducting gap is smaller than the pseudogap, as depicted by the white arrows.
This competition between pseudo- and superconducting gaps is consistent with other cluster DMFT studies\cite{civelli08,sordi12,gull12}. 

In conclusion, by combining Raman experiment and CDMFT, we have explored the unoccupied part of the single-particle spectra of an underdoped cuprate and found that the pseudogap opens {\it above} the Fermi level in the nodal region. The pseudogap thus shows a strongly electron-hole asymmetric $s$-wave structure, distinct from the $d$-wave superconducting gap. This suggests that they have different origins.
To obtain this result, it has been crucial to shed light on the empty part (dark side) of the electronic spectrum. This region should therefore be the focus of future experimental (e.g., along the lines of Refs.~\onlinecite{kanigel08,yang08,hashimoto10,he11,hoffman10}) and theoretical (as focused in Refs.~\onlinecite{sakai10,yamaji11}) developments.

We acknowledge valuable comments by A. Tremblay. M. Ci. acknowledges discussions with V. Brouet, A. Cano, B. G. Kotliar, I. Paul, and A. Santander-Syro.
The work was supported by a Grant-in-Aid for Scientific Research (Grant No. 22340090), from MEXT, Japan. A part of the research has been funded by the Strategic Programs for Innovative Research (SPIRE), MEXT, and the Computational Materials Science Initiative (CMSI), Japan. K.H. is supported by the Austrian Science Fund (FWF) through SFB ViCoM F4103-N13, and G.S. by the FWF under ``Lise-Meitner" Grant No. M1136. S.B., Y.G., M.Ca, M.-A.M., and A.S. acknowledge support from Agence Nationale de la Recherche through Grant BLAN07-1-183876, ``GAPSUPRA". The work in BNL is supported by the DOE under Contract No. DE-AC02-98CH10886. The calculations were performed at the Vienna Scientific Cluster and at the Supercomputer Center, ISSP, University of Tokyo.

\clearpage

\begin{center}
 \bf \LARGE Supplementary Information
\end{center}

\section*{I. Comparison of energy scales}
Although the aim of our theory is not in a quantitative description of real materials, our results are still in reasonable quantitative agreements with experimental results on cuprates.
We summarize in the Table below the comparison of various energies and temperatures between our theory and Raman experiment.
We employ $t$=0.3eV, which is a typical value for cuprates\cite{andersen95}.
In order to take account of the fact that the pseudogap amplitude is about 1.5 times larger at 5\% doping (where the theoretical calculation was done) than at 11\% doping (where the experiment was done), we simply divide the theoretical energy scales by a factor of 1.5, and show that we can obtain reasonable estimates at 11\% doping.

\begin{table}[h] 
\begin{center} 
\begin{tabular}{cccc} \hline\\ [-15pt]
  & Theory (5\%) \ \ \ & Estimate at 11\% \ \ \ & Experiment (11\%)\\ 
\hline \\ [-15pt]
$T^\ast_\text{B2g}$ & 0.08$t$   & 0.053$t$$\sim$180K & 150K \\
$T^\ast_\text{B1g}$ & 0.12$t$ & 0.08$t$$\sim$280K & 200K \\
$\D_\text{B2g}$ & 0.3$t$    & 0.2$t$$\sim$480cm$^{-1}$ & 450cm$^{-1}$\\
$\D_\text{B1g}$ & 0.35$t$   & 0.23$t$$\sim$560cm$^{-1}$ & 550cm$^{-1}$\\
\hline\\
\end{tabular} 
\begin{minipage}{0.48\textwidth}
\begin{flushleft}
\textbf{Table:} Comparison of various energy scales between our theory and Raman experiment. Estimate at 11\% doping is obtained by simply dividing the theoretical value at 5\% doping by a factor of 1.5 and using $t$=0.3eV.
$\D_\text{B2g,B1g}$ is the energy interval where the depression due to pseudogap occurs in Figs.~2a-d.
\end{flushleft}
\end{minipage}
\end{center} 
\end{table}

\vspace{-20pt}
\section*{II. Detailed comparisons with ARPES, STM, Raman, and optics experiments}

In Fig.~S1 the theoretical spectral intensity at 5\% doping (Figs.~1e-g in the main text) is compared with the ARPES experimental result\cite{yang08} by Yang {\it et al.} for underdoped Bi2212 ($T_\text{c}$=65K, $p$$\sim$0.09).
Here the theoretical energy scale is once again fixed at $t$=0.3eV, which, as we have shown above, looks appropriate for the comparison with experiments (upon considering the doping difference between experiment and theory). We stress however that the comparison in Fig.~S1 must be considered under the mere qualitative point of view, as in general we do not expect the quasiparticle dispersion to scale with doping as the pseudogap amplitude.

Since the pseudogap amplitude at 9\% doping is about 30\% smaller than that at 5\% doping, the ingap state seen in Fig.~1g at $\w\gtrsim 0.4t$ should be situated around 0.28$t$$\sim$0.08eV at 9\% doping, which is beyond the experimental window ($\lesssim$0.04eV) of Fig.~S1g. 

\begin{figure}[h]
\begin{center}
\includegraphics[width=0.48\textwidth]{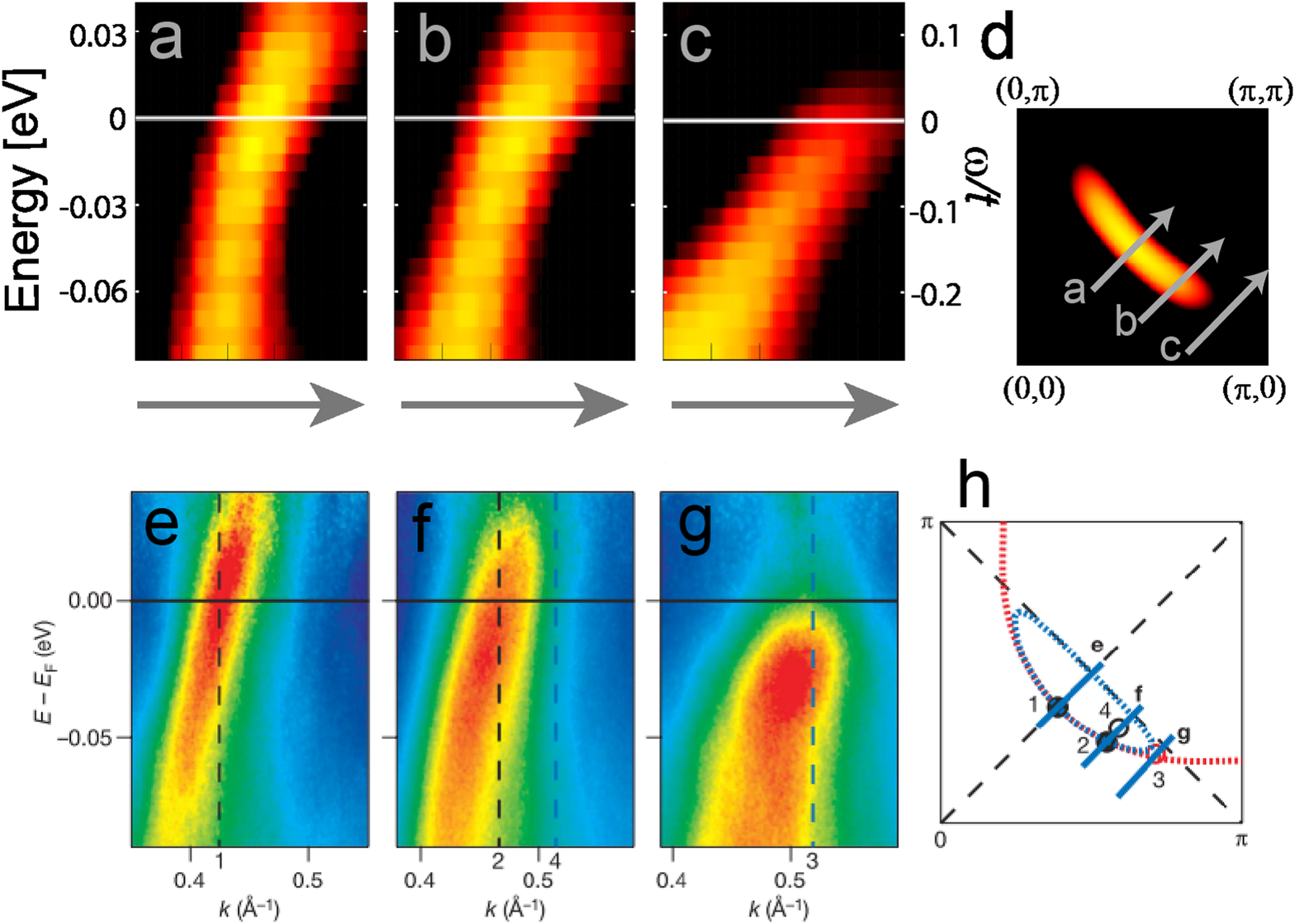}
\begin{minipage}{0.48\textwidth}
\begin{flushleft}
\vspace{10pt}
\textbf{Fig.~S1:} Comparison of the one-particle spectra between our theory (a-d) and the ARPES experiment (e-h) of Ref.~\onlinecite{yang08}.
The energy scale evaluated for $t$=0.3eV is put on the left hand side of a.
\end{flushleft}
\end{minipage}
\end{center}
\end{figure}

In Fig.~S2 we compare our theoretical $dI/dV$ curve at 5\% doping with the STM result\cite{pushp09} by Pushp {\it et al.} for a very strongly underdoped Bi2212 sample ($T_\text{c}$=35K, $p$$\sim$0.07).
Though the temperature ($T$$=$0.06$t$) in theory (Fig.~S2a) would be substantially higher than the highest temperature (71K) in Fig.~S2b, we can still find various similarities in the two results:
i) The slope on the occupied side is longer than that on the unoccupied side, showing a strong electron-hole asymmetry.
ii) The minimum is slightly above $V$=0. This is also seen in Fig.~4a of Ref.~\onlinecite{gomes07} (reproduced in Fig.~S2c) where the minimum of UD73 sample is shifted to positive energy, compared to that of UD83 sample. 
A similar shift can also be seen in Fig.~3c of Ref.~\onlinecite{Lee09}.
iii) The pseudogap ends around 100mV (which is still in a reasonable agreement even if we consider the doping variation of the pseudogap amplitude), and the spectrum saturates above this energy.

\begin{figure*}[t]
\begin{center}
\includegraphics[width=0.85\textwidth]{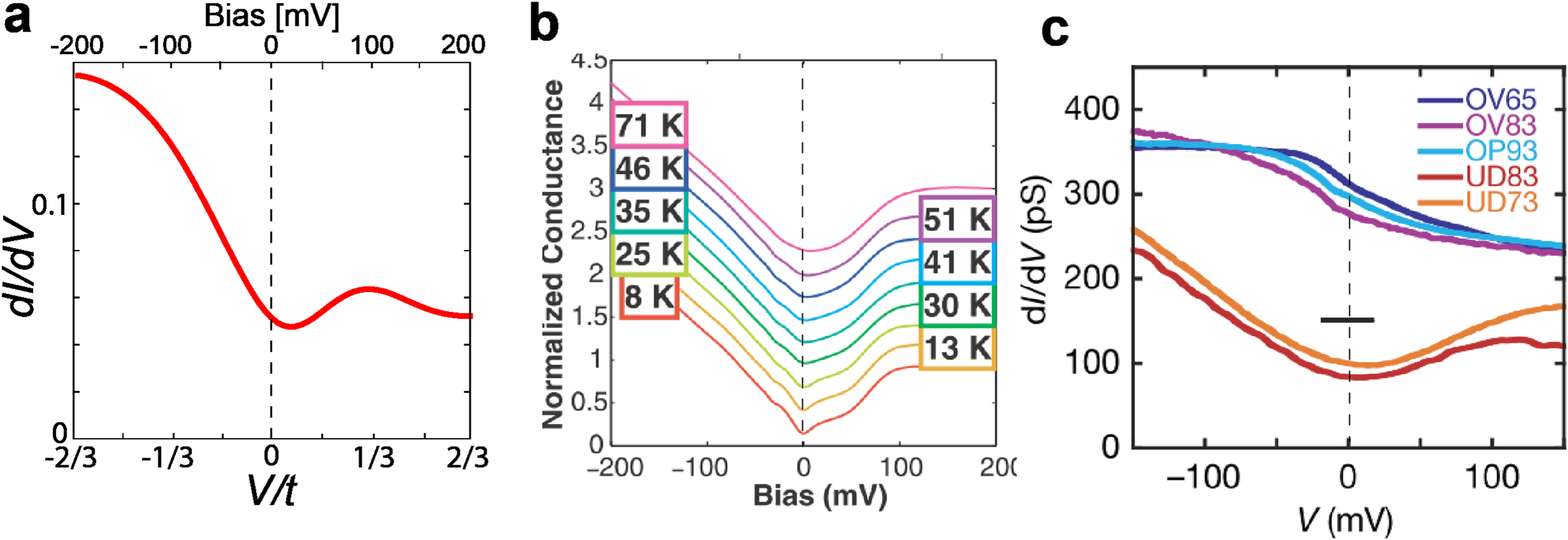}
\begin{minipage}{0.95\textwidth}
\begin{flushleft}
\vspace{10pt}
\textbf{Fig.~S2:} Comparison of ({\bf a}) calculated tunneling conductance with ({\bf b}) that of STM experiment (after Fig.~5B inset in Ref.~\onlinecite{pushp09}) for a strongly underdoped Bi2212 sample with $T_\text{c}$=35K.
The bias voltage evaluated for $t$=0.3eV is put on the top of the left panel to facilitate a comparison with the experiment.
{\bf c}, STM experimental result (Fig.~4a in Ref.~\onlinecite{gomes07}) for various Bi2212 samples at $T$$>$$T_\text{c}$.
\end{flushleft}
\end{minipage}
\end{center}
\end{figure*}

\begin{figure}[!]
\begin{center}
\includegraphics[width=0.45\textwidth]{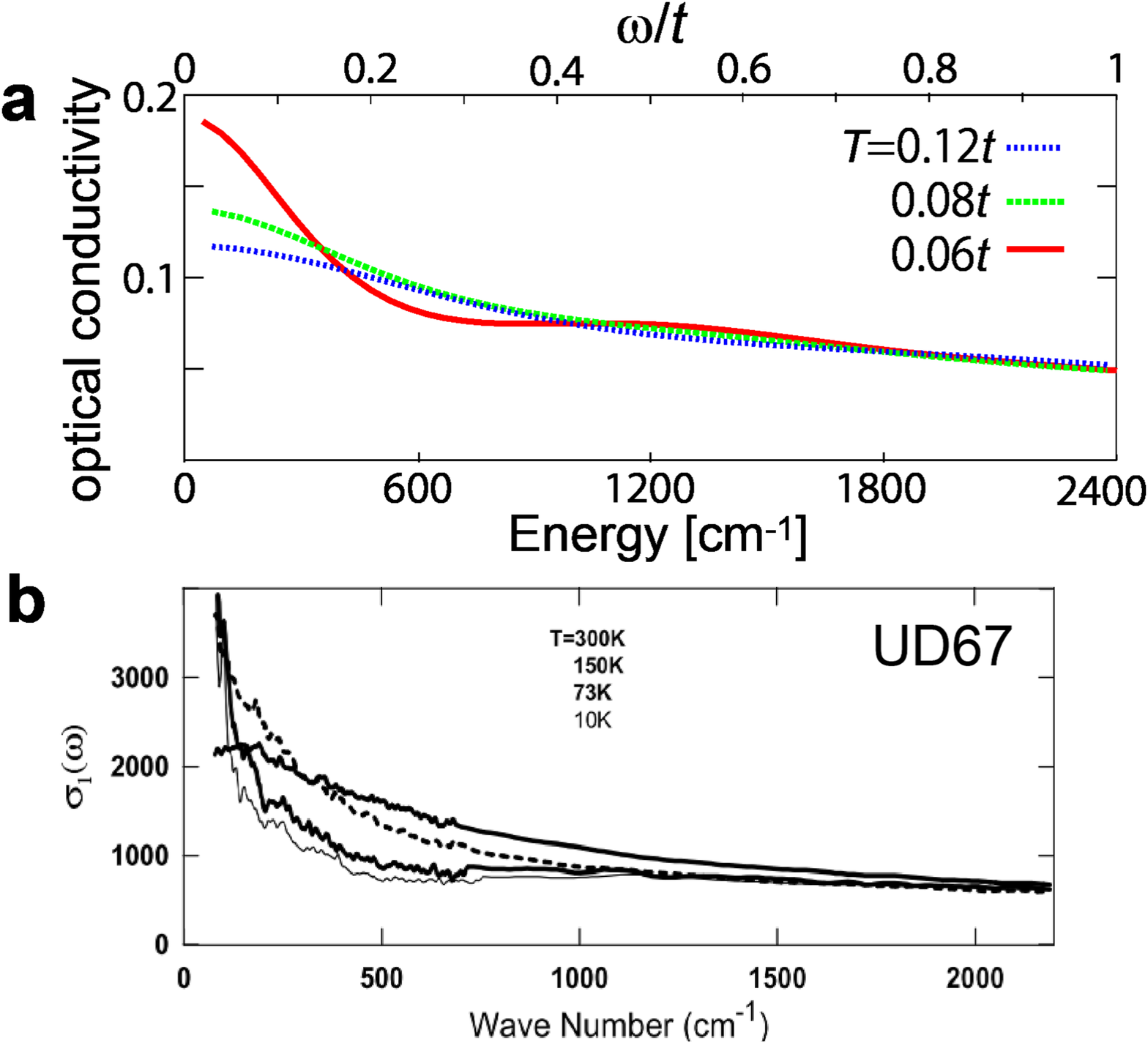}
\begin{minipage}{0.48\textwidth}
\begin{flushleft}
\vspace{10pt}
\textbf{Fig.~S3:} Comparison of ({\bf a}) calculated optical conductivity with ({\bf b}) that of experiment (after Fig.~6 of Ref.~\onlinecite{puchkov96}) for a strongly underdoped Bi2212 sample with $T_\text{c}$=67K.
The energy scale evaluated for $t$=0.3eV is put on the bottom of a to facilitate a comparison with the experiment.
\end{flushleft}
\end{minipage}
\end{center}
\end{figure}

\begin{figure}[!]
\begin{center}
\includegraphics[width=0.3\textwidth]{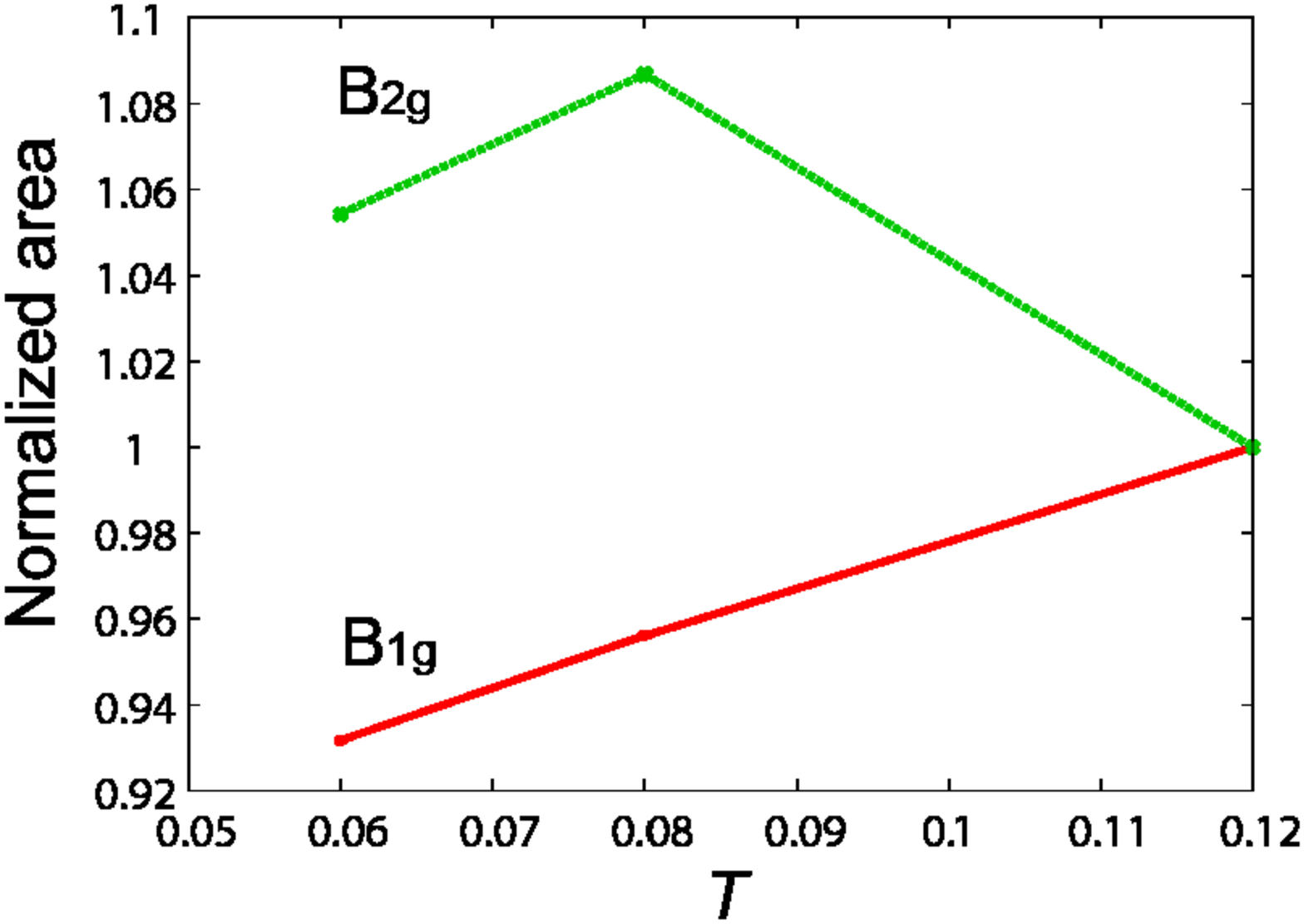}
\begin{minipage}{0.48\textwidth}
\begin{flushleft}
\vspace{10pt}
\textbf{Fig.~S4:} Temperature dependence of the area under the theoretical Raman curves of Figs.~2(b) and 2(d), normalized at $T$$=$$0.12t$.
The cutoff energy is set to be $0.6t$. The figure is comparable with the insets of Figs.~2(a) and 2(c).
\end{flushleft}
\end{minipage}
\end{center}
\end{figure}

Figure S3 compares the optical conductivity calculated by the CDMFT with the experimental result\cite{puchkov96} by Puchkov {\it et al.} for a strongly underdoped Bi2212 sample ($T_\text{c}$=67K, $p$$\sim$0.09).
We see in both experiment and theory that 
i) at high temperatures the conductivity is rather flat, 
ii) as $T$ is lowered, the low-energy part rapidly increases, and 
iii) as the pseudogap opens, the conductivity decreases a lot in an intermediate-energy region (around 0.25$t$ in theory and 400cm$^{-1}$ in experiment) in a fashion similar to our B$_\text{2g}$ Raman response.
This consistency with optical experiments also supports our results.

Figure S4 shows the integrated area under the theoretical Raman response curves in Fig.~2(b) and 2(d), normalized at $T$$=$$0.12t$$\sim$$T_\text{B1g}^\ast$. The cutoff energy of the integration is set to be $0.6t$. As temperature decreases, the area of B$_{1g}$ monotonically decreases while that of B$_{2g}$ shows a non-monotonic increase-decrease behavior, in accordance with the experimental results in the insets of Fig.~2(a) and 2(c).

\section*{III. Relation to previous theories}
The momentum structure shown in Fig.~3 is qualitatively consistent with the previous exact diagonalization (ED) result\cite{tohyama04} for the two-dimensional (2D) $t$-$J$ model, and with the 2$\times$2 CDMFT+ED result\cite{sakai09,kyung06} for the 2D Hubbard model, as well as with a recent extended slave-boson mean-field theory\cite{yamaji11}. In this latter theory, in particular, the pseudogap is identified with a hybridization gap between the quasiparticle and a composite-fermion excitation, which is an electron trapped by a hole. This pseudogap shows also an $s$-wave form and is situated above the Fermi level around the nodal region, similarly to the CDMFT results. 
In previous 16-site dynamical cluster approximation (DCA) studies\cite{macridin06}, performed at $T$$=$0.12$t$, the antinodal pseudogap was observed, but the temperature was too high to observe the pseudogap in the nodal region. A recent 8-site DCA study at a comparably low temperature ($T$$=$0.05$t$ for $U$$=$7$t$) shows a small suppression in the low-energy spectrum around the node at a low doping\cite{lin10}. The consistency between several different theories further supports our $s$-wave pseudogap interpretation of Raman experiments. 

\section*{IV. Robustness of $s$-wave pseudogap structure against the choice of periodization scheme}

\begin{figure}[h]
\begin{center}
\includegraphics[width=0.4\textwidth]{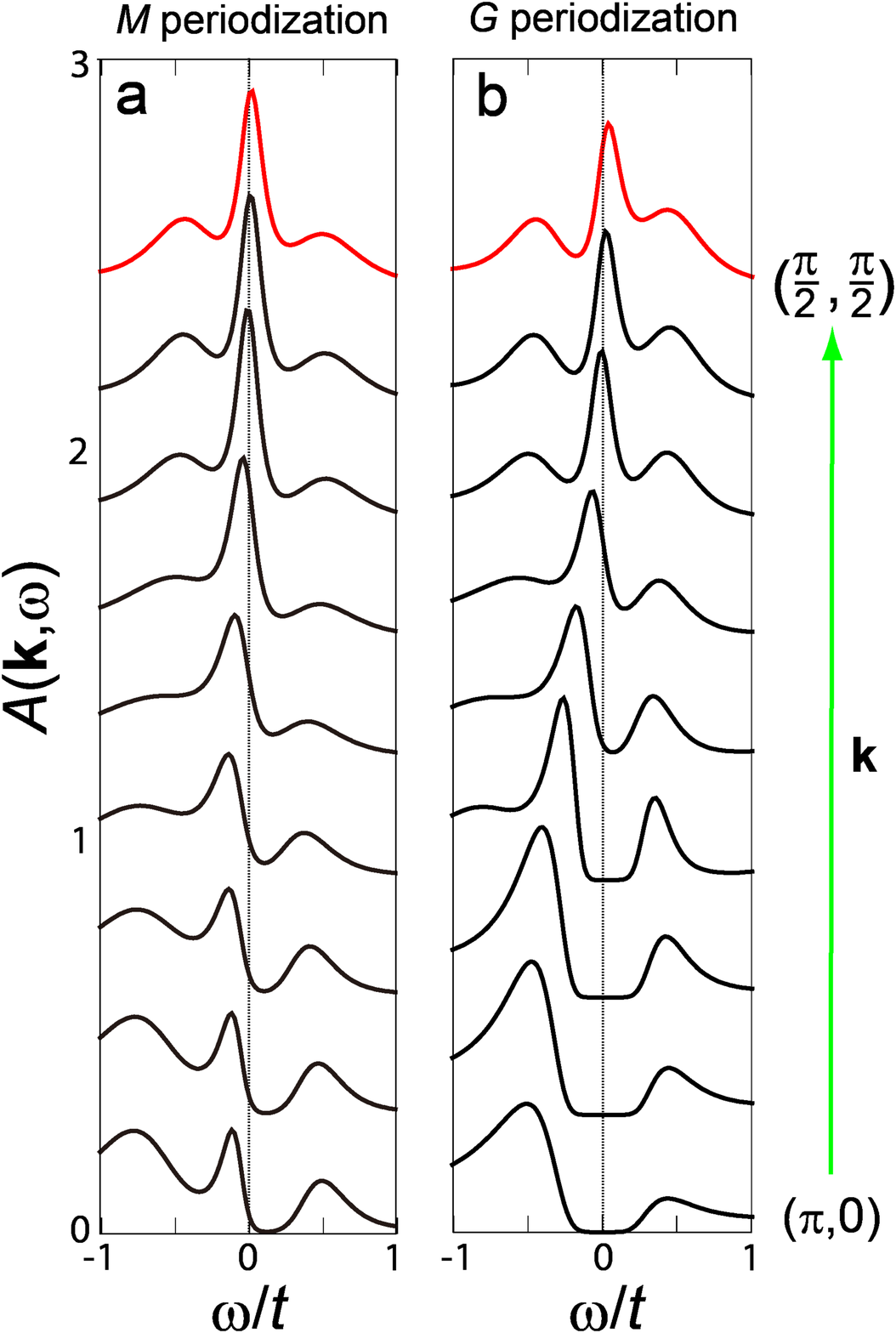}
\begin{minipage}{0.48\textwidth}
\textbf{Fig.~S5:} Spectral functions obtained by $M$ and $G$ periodizations for $T=0.06t$.
\end{minipage}
\end{center}
\end{figure}

\begin{figure}[h]
\begin{center}
\includegraphics[width=0.45\textwidth]{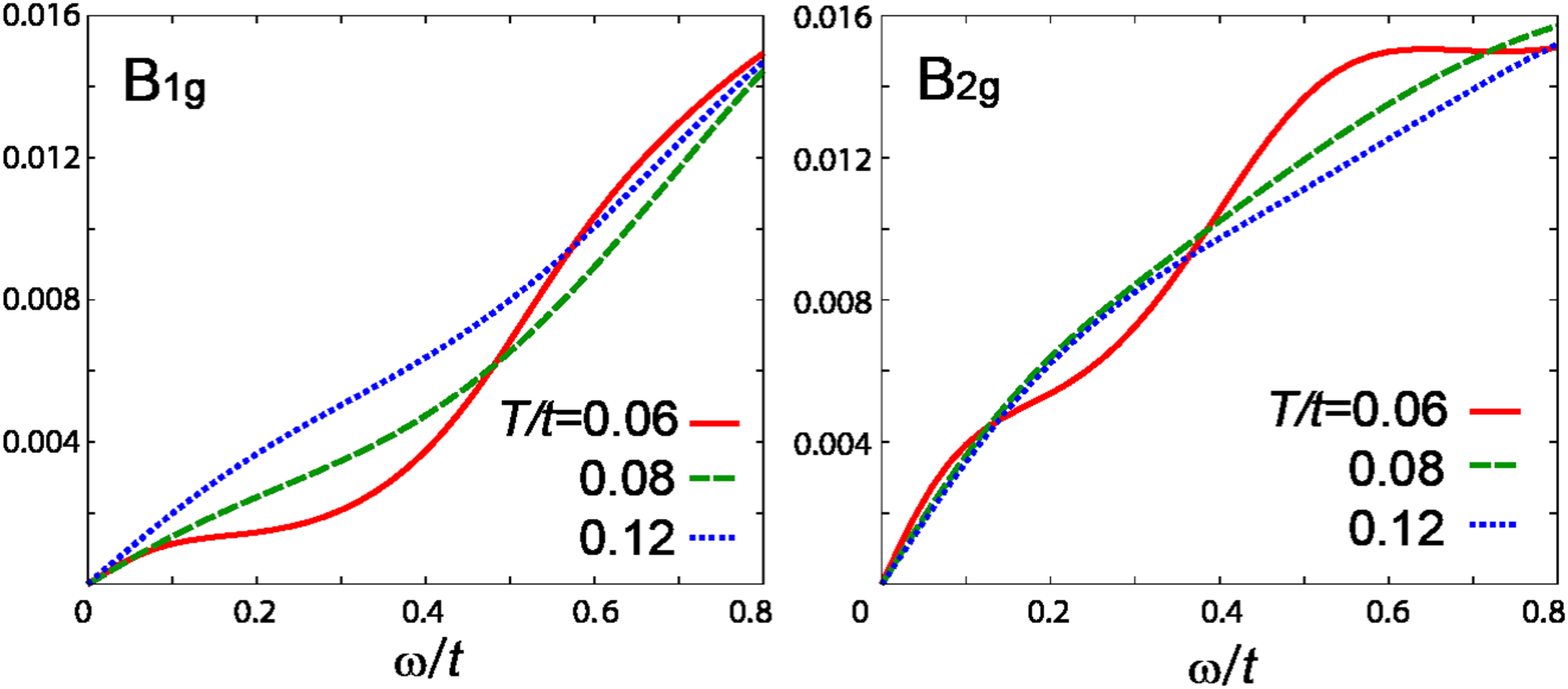}
\begin{minipage}{0.48\textwidth}
\textbf{Fig.~S6:} $B_\text{1g}$ and $B_\text{2g}$ Raman spectra calculated from the $G$-periodized Green's function.
\end{minipage}
\end{center}
\end{figure}

\begin{figure}[h]
\begin{center}
\includegraphics[width=0.3\textwidth]{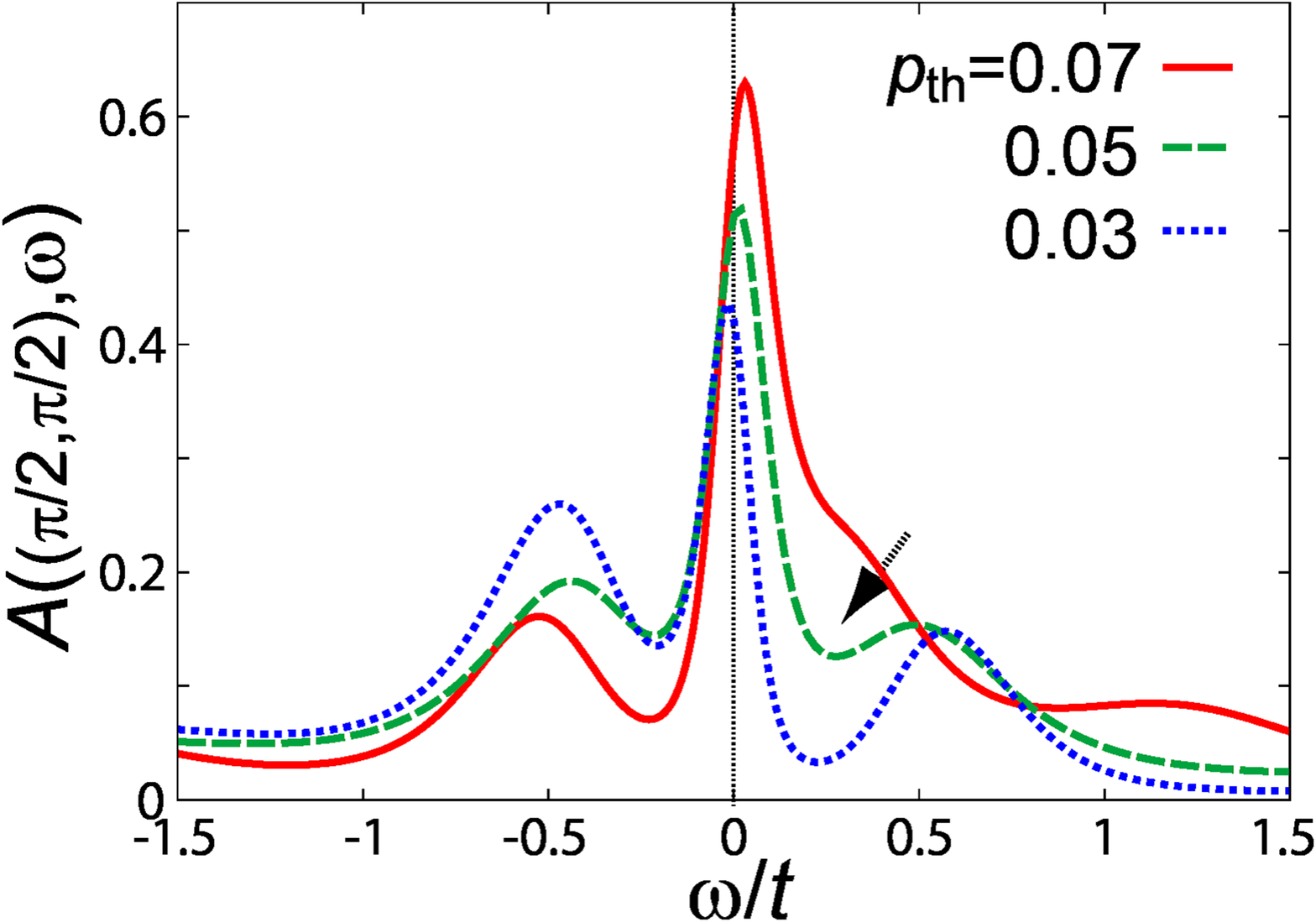}
\begin{minipage}{0.48\textwidth}
\begin{flushleft}
\textbf{Fig.~S7:} Theoretical one-particle spectra at the node for various dopings $p_\text{th}$ at $T$$=$0.06$t$.
\end{flushleft}
\end{minipage}
\end{center}
\label{fig:ndep}
\end{figure}

In the cellular DMFT\cite{kotliar01}, a momentum-dependent function such as the single-particle spectral function $A(\Vec{k},\w)$ is calculated through a periodization scheme, which is a Fourier transformation truncated at the cluster size.
In the present work we employ the cumulant ($M$) periodization\cite{stanescu06}, since it gives the fastest convergence against the cluster size for the parameter region we studied \cite{sakai12}.

However, we may alternatively employ the Green's function ($G$) periodization\cite{kyung06}, which can also give a reasonable result.
Since the $G$ periodization can give slightly different spectra from those of the $M$ periodization, it is worthwhile checking if our conclusion is robust against the choice of periodization scheme.
Figure S5 compares the two spectral functions obtained by $M$ and $G$ periodizations, respectively. While the pseudogap in the antinodal region is larger and more symmetric in $G$ scheme than in $M$ scheme, the spectra in the nodal region share essentially the same structure (as expected from Fig.~6 in Ref.~\onlinecite{sakai12}, which shows a good agreement of low-energy selfenergy around the node) so that $s$-wave pseudogap is common to both schemes.
The main gap position at $\omega$$\sim$0.2$t$ inferred from Fig.~1a and from Fig.~S5a would be corrected much closer to $\omega$=0 when the G periodization is employed as is inferred from Fig.~S5b. The $dI/dV$ curve shown in Fig.~1h is calculated directly from the cluster density of states without possible small errors coming from periodization procedure.
This cluster density of states coincides with the angle integration of the $G$-periodized spectra shown in Fig.~S5b.
The slight uncertainty of the size and the position of the gap in the antinodal region, depending on the choice of the periodization scheme, does not alter our main conclusion about the $s$-wave-like structure of the pseudogap.

The Raman responses calculated from the $G$-periodized Green's function are plotted in Fig.~S6. The figure shares essential features with $M$-periodized results [Fig.~2(b) and (d)]: As $T$ decreases, (i) B$_\text{1g}$ response shows a monotonic decrease (indicating gap opening) while B$_\text{2g}$ response shows (ii) a monotonic increase at low energy (indicating metallic property), and (iii) a non-monotonic change in intermediate-energy (indicating gap opening at finite energy). The larger depression in B$_\text{1g}$ response, compared with Fig.~2(d), would be attributed to the larger antinodal gap seen in $G$-periodized one-electron spectra (Fig.~S5b).

\section*{V. Doping dependence of the single-particle spectra}

We compare in Fig.~S7 the theoretical spectral function $A(\Vec{k},\w)$ at the node for various dopings $p_\text{th}$ at $T$$=$0.06$t$. We have employed an $N_\text{C}$$=$4$\times$3 cluster only for $p_\text{th}$$=$0.07 because of a severe sign problem for $N_\text{C}$$=$16. The pseudogap in the nodal region becomes more distinct at lower doping. This is analogous to the well-established behavior of the antinodal pseudogap\cite{sakai10,kyung06,stanescu06,liebsch09,lin10,macridin06,ferrero09}. Combining with Fig.~3, it is likely that $T^\ast_\text{N}$ decreases with doping as is the case of $T^\ast_\text{AN}$. 
Note that the large reduction around $\w$$<$$-t$ would correspond to the waterfall\cite{graf07,valla07,meevasana07}, observed in high-$T_\text{c}$ cuprates while a small dip just below the quasiparticle peak would correspond to the low-energy kink\cite{sakai10}.


\begin{references}
\item[\nonumber]
\bibitem{tsuei97}
C. C. Tsuei {\it et al}., Nature {\bf 387}, 481 (1997).
\bibitem{norman05}
M. R. Norman, D. Pines, and C. Kallin, Adv. Phys. {\bf 54}, 715 (2005).
\bibitem{hufner08}
S. H\"ufner, M. A. Hossain, A. Damascelli, and G. A. Sawatzky, Rep. Prog. Phys. {\bf 71}, 062501 (2008); M. Le Tacon, A. Sacuto, A. Georges, G. Kotliar, Y. Gallais, D. Colson, and A. Forget, Nat. Phys. 2, 537 (2006). 
\bibitem{millis06}
A. J. Millis, Science {\bf 314}, 1888 (2006).
\bibitem{kanigel08}
A. Kanigel {\it et al}., Phys. Rev. Lett. {\bf 101}, 137002 (2008). 
\bibitem{yang08}
H.-B. Yang {\it et al}., Nature {\bf 456}, 77 (2008).
\bibitem{hashimoto10}
M. Hashimoto {\it et al}., Nature Phys. {\bf 6}, 414 (2010).
\bibitem{he11}
R.-H. He {\it et al}., Science {\bf 331}, 1579 (2011).
\bibitem{hoffman10}
J. E. Hoffman, Nature Phys. {\bf 6}, 404 (2010).
\bibitem{anderson87}
P. W. Anderson, Science {\bf 235}, 1196 (1987).
\bibitem{emery95}
V. J. Emery, and S. A. Kivelson, Nature (London) {\bf 374}, 434 (1995).
\bibitem{chakravarty01}
S. Chakravarty, R. B. Laughlin, D. K. Morr, and C. Nayak, Phys. Rev. B {\bf 63}, 094503 (2001).
\bibitem{balents98}
L. Balents, M. P. A. Fisher, and C. Nayak, Int. J. Mod. Phys. B {\bf 12}, 1033 (1998).
\bibitem{franz01}
M. Franz, and Z. Tesanovic, Phys. Rev. Lett. {\bf 87}, 257003 (2001).
\bibitem{yang06}
K.-Y. Yang, T. M. Rice, and F.-C. Zhang, Phys. Rev. B {\bf 73}, 174501 (2006).
\bibitem{damascelli03}
A. Damascelli, Z. Hussain, and Z.-X. Shen, Rev. Mod. Phys. {\bf 75}, 473 (2003).
\bibitem{kotliar01} 
G. Kotliar, S. Y. Savrasov, G. Palsson, and G. Biroli, Phys. Rev. Lett. {\bf 87}, 186401 (2001).
\bibitem{doiron07}
N. Doiron-Leyraud {\it et al.}, Nature {\bf 447}, 565 (2007).
\bibitem{sebastian08}
S. E. Sebastian {\it et al.}, Nature {\bf 454}, 200 (2008).
\bibitem{devereaux94}
T. P. Devereaux {\it et al.}, Phys. Rev. Lett. {\bf 72}, 396 (1994).
\bibitem{sakai12}
S. Sakai {\it et al}., Phys. Rev. B {\bf 85}, 035102 (2012).
\bibitem{kyung06}
B. Kyung {\it et al}., Phys. Rev. B {\bf 73}, 165114 (2006).
\bibitem{sakai09}
S. Sakai, Y. Motome, and M. Imada, Phys. Rev. Lett. {\bf 102}, 056404 (2009).
\bibitem{sakai10}
S. Sakai, Y. Motome, and M. Imada, Phys. Rev. B {\bf 82}, 134505 (2010).
\bibitem{senechal04}
D. S\'en\'echal, and A.-M. S. Tremblay, Phys. Rev. Lett. {\bf 92}, 126401 (2004).
\bibitem{civelli05}
M. Civelli, M. Capone, S. S. Kancharla, O. Parcollet, and G. Kotliar, Phys. Rev. Lett. {\bf 95}, 106402 (2005).
\bibitem{stanescu06}
T. D. Stanescu and G. Kotliar, Phys. Rev. B {\bf 74}, 125110 (2006).
\bibitem{liebsch09}
A. Liebsch and N.-H. Tong, Phys. Rev. B {\bf 80}, 165126 (2009).
\bibitem{civelli08}
M. Civelli {\it et al}., Phys. Rev. Lett. {\bf 100}, 046402 (2008).
\bibitem{civelli09}
M. Civelli, Phys. Rev. B {\bf 79}, 195113 (2009).
\bibitem{gull11}
E. Gull {\it et al}., Rev. Mod. Phys. {\bf 83}, 349 (2011).
\bibitem{andersen95}
O. K. Andersen, A. I. Liechtenstein, O. Jepsen, and F. Paulsen, J. Phys. Chem. Solids {\bf 56}, 1573 (1995).
\bibitem{devereaux07}
T. P. Devereaux and R. Hackl, Rev. Mod. Phys. {\bf 79}, 175 (2007).
\bibitem{tohyama04}
T. Tohyama, Phys. Rev. B {\bf 70}, 174517 (2004).
\bibitem{berthod06}
C. Berthod, T. Giamarchi, S. Biermann, and A. Georges, Phys. Rev. Lett. {\bf 97}, 136401 (2006).
\bibitem{yamaji11}
Y. Yamaji and M. Imada, Phys. Rev. Lett. {\bf 106}, 016404 (2011).
\bibitem{anderson-ong06}
P. W. Anderson and N. P. Ong, J. Phys. Chem. Solids {\bf 67}, 1 (2006); P. W. Anderson, Nat. Phys. {\bf 2}, 626 (2006).
\bibitem{pushp09}
A. Pushp {\it et al.}, Science {\bf 324}, 1689 (2009).
\bibitem{devereaux99}
T. P. Devereaux and A. P. Kampf, Phys. Rev. B {\bf 59}, 6411 (1999).
\bibitem{lin12}
N. Lin, E. Gull, and A. J. Millis, Phys. Rev. Lett. {\bf 109}, 106401 (2012).
\bibitem{maier05}
T. Maier, M. Jarrell, T. Pruschke, and M. H. Hettler, Rev. Mod. Phys. {\bf 77}, 1027 (2005).
\bibitem{phonon}
Note that sharp peaks (e.g., at $\sim$120cm$^{-1}$ in B$_\text{2g}$, and at $\sim$100 and $\sim$280cm$^{-1}$ in B$_\text{1g}$) in Figs.~2(a) and (c) stem from phonon excitations, which we will not consider in our study.
\bibitem{macridin06}
A. Macridin, M. Jarrell, T. Maier, P. R. C. Kent, and E. D'Azevedo, Phys. Rev. Lett. {\bf 97}, 036401 (2006).
\bibitem{venturini02}
F. Venturini {\it et al}., Phys. Rev. Lett. {\bf 89}, 107003 (2002).
\bibitem{opel00}
M. Opel {\it et al}., Phys. Rev. B {\bf 61}, 9752 (2000).
\bibitem{gallais05}
Y. Gallais, A. Sacuto, T. P. Devereaux, and D. Colson, Phys. Rev. B {\bf 71}, 012506 (2005).
\bibitem{ferrero09}
M. Ferrero {\it et al}., Phys. Rev. B {\bf 80}, 064501 (2009).
\bibitem{lin10}
N. Lin, E. Gull, and A. J. Millis, Phys. Rev. B {\bf 82}, 045104 (2010).
\bibitem{nemetschek97}
R. Nemetschek {\it et al}., Phys. Rev. Lett. {\bf 78}, 4837 (1997).
\bibitem{sordi12}
G. Sordi, P. S\'emon, K. Haule, and A.-M. S. Tremblay, Phys. Rev. Lett. {\bf 108}, 216401 (2012).
\bibitem{gull12}
E. Gull, O. Parcollet, and A. J. Millis, Phys. Rev. Lett. {\bf 110}, 216405 (2013).
\end{references}

\begin{references}
\bibitem[52]{gomes07}
K. K. Gomes {\it et al.}, Nature {\bf 447}, 569 (2007).
\bibitem[53]{Lee09}
J. Lee {\it et al.}, Science {\bf 325}, 1099 (2009).
\bibitem[54]{puchkov96}
A. V. Puchkov, D. N. Basov, and T. Timusk,
J. Phys.: Condens. Matter {\bf 8}, 10049 (1996).
\bibitem[55]{graf07}
J. Graf {\it et al}., Phys. Rev. Lett. {\bf 98}, 067004 (2007).
\bibitem[56]{valla07}
T. Valla {\it et al}., Phys. Rev. Lett. {\bf 98}, 167003 (2007).
\bibitem[57]{meevasana07}
W. Meevasana {\it et al}., Phys. Rev. B {\bf 75}, 174506 (2007).
\end{references}
\end{document}